# Fast Mid-IR Flashes Detected During Small Solar X-Ray Bursts

Marta M. Cassiano[1], Pierre Kaufmann[1,2], Rogério Marcon[3,4], Amauri S. Kudaka[1], Adolfo Marun[5], Rodolfo Godoy[5], Pablo Pereyra[5], Arline M. Melo[1], Hugo Levato[6]

[1] Centro de Radioastronomia e Astrofísica Mackenzie, Escola de Engenharia, Universidade Presbiteriana Mackenzie, São Paulo, Brazil
[2] Centro de Componentes Semicondutores, Universidade Estadual de Campinas, Campinas, Brazil
[3] Instituto de Física Gleb Wataghin, Universidade Estadual de Campinas, Campinas, Brazil
[4] Observatório Solar Bernard Lyot, Campinas, Brazil
[5] Complejo Astronómico El Leoncito, CONICET, San Juan, Argentina
[6] Instituto de Ciencias Atmosféricas y del Espacio, CONICET, San Juan, Argentina

**Abstract**

Solar observations in the mid-infrared 8-14 µm band continuum were carried out with cadence of 5 frames per second, in December 2007. Rapid small heated sources, with typical duration of the order of seconds, were found on the bright plage-like areas around sunspots, in association with relatively weak GOES soft X-ray bursts. This work presents the analysis of fast mid-infrared flashes detected during a GOES B2.0-class event on 10 December 2007, beginning at about 10:40 UT. Rapid brightness temperature enhancements of 0.5 to 2.0 K were detected at the Earth by a microbolometer array, using a telescope with 10.5 cm diameter aperture producing a diffraction limited field-of-view of 25 arcsec. Minimum detectable temperature change was of 0.1 K. The corresponding fluxes are 30-130 solar flux units. At the solar surface the estimated rapid brightenings were of 50-150 K.

1. Introduction

The first mid-infrared solar observations using a single detector photometer coupled to a two-mirror scanner device at the 150 cm MacMath Solar Telescope, Kitt Peak, AZ, have revealed a network of small bright areas in the vicinity of sunspots (Turon and Léna, 1970). Hudson and Lindsey (1974) and Hudson (1975) observed active regions at 20 µm and 350 µm using single photometers at the 152 cm Mt. Lemmon telescope, AZ. 300 seconds oscillations were identified at 20 µm and upper limits were set for flare detectability. Gezari, Livingston, and Varosi (1999) carried out the first program of observations with a mid-infrared two-dimensional 58 x 62 array camera system. The images showed bright plage-like structures around sunspots, not observed in continuum visible light images but resembling structures seen in Ca K images.

The renewed interest to observe solar activity in the far- and near-infrared range arose from the detection of solar bursts at submillimeter wavelengths indicating a new spectral component, with flux density increasing with the frequency towards the THz range (i.e., for frequencies larger than about 0.1 THz), distinct from the well-known microwave component (where fluxes usually exhibit maxima in the range 5-20 GHz) (Kaufmann *et al.,* 2004; 2009; Silva *et al.*, 2007; Cristiani *et al.,* 2008). These results brought difficulties for the interpretation of flaring physical processes, pointing to the need of a complete spectral description to be obtained with observations at the infrared band.

New setups for ground-based solar measurements at 10 μm were developed (Melo *et al.*, 2006; Marcon *et al.*, 2008). The first observations confirmed the presence of mid-infrared bright plage-like areas around sunspots, associated with the regions observed in Ca II line emissions and with magnetic structures. These observations also revealed variations with time scales of tens of seconds during a solar flare.

A flare detection campaign was carried out on a longer period, adding considerable improvements in data acquisition hardware and on image processing procedures. Several solar events were recorded with high cadences in the period of 8-13 December 2007. Here we show in detail the results for the 10 December 2007 event associated to the GOES B2.0-class burst beginning at about 10:40 UT.

## 2. Instrumentation and Data Reduction

The December 2007 mid-infrared solar observations were carried out at El Leoncito Astronomical Complex in San Juan, Argentina, using the same optical setup previously described (Melo *et al.*, 2006). It consisted of a 30 cm flat mirrors coelostat followed by an arrangement composed of a 10.5 cm concave spherical mirror aperture, a convex mirror, and a concave mirror to allow the afocal alignment to produce the image at the microbolometer camera. A germanium flat disk followed by a lens precede the Wuhan Guide IR928 microbolometer camera, with a focal plane array containing 320 x 240 uncooled amorphous silicon detectors. The diffraction-limited field-of-view was of 25 arcsec, defined as ~ 70 $\lambda/D$ deg, where $\lambda$ is the wavelength and D the aperture diameter (Kraus, 1986).

The images were captured at a rate of 5 frames per second by a Euresys GrabLink board. The data reduction was done using the IDL software - *Interactive Data Language*. A set of new computational routines were developed intended for image treatment for the mid-infrared solar data application.

Figure 1 shows an example of an original full disk frame captured during the observations of 10 December 2007. The work area is the larger square on the disk, with dimensions of 80 x 80 pixels (14.7 x 14.7 arcmin on the Sun). It contains the selected 25 x 25 pixels box (4.6 x 4.6 arcmin) where the active region is located, upper left. The following procedures were sequentially applied for the treatment of the selected portion of the image:

a) Running mean of 10 frames, smoothing the time profile and keeping the same sampling rate (5 per second).
b) A flat fielding procedure was applied (Marcon *et al.,* 2008).
c) The value in every pixel was replaced by the averaged value over a region of 3 x 3 pixels (i.e., 33.2 x 33.2 arcsec), slightly larger than the diffraction limited field-of-view.
d) During the observations, the camera performs periodic and automatic calibrations of temperature and pixels gains ("internal flat field"). For this reason, the frames present, sometimes, variations in the overall level. To compensate the offsets caused by such variations, for every frame we subtracted the 25 x 25 pixels box containing the flaring region (upper left box in Figure 1) from the mean value calculated over all the auxiliary 25 x 25 pixels box located away from the active region (lower right box in Figure 1). This method of offset correction should ensure that the variations represent, in fact, real solar variations.

Figure 1 also shows a dashed 25 x 25 pixels control box for comparing the image stability for a neighboring quiescent region.

Figure 2 presents a resulting image (4.6 x 4.6 arcmin) after application of the procedures described above. It shows the sunspots (dark areas, labeled A and B) and the plages (bright regions) around them. Spot C also can be distinguished with less contrast. Note that the position angle of this figure is rotated clockwise by about 35$^o$ with respect to the solar N-S direction.

## 3. Solar Activity at the 10 μm Band

The mid-infrared camera used in the observations has an internal system of temperature control that displays two crosses and provides temperature readings of specific pixels at the image. One of the crosses remains at a fixed position, reading a reference temperature ($T_R$). The second cross moves indicating the temperature of the hottest pixel in the whole field ($T_H$). Both $T_R$ and $T_H$ are printed on every frame recorded and the temperature difference ($\Delta T = T_H - T_R$) between both pixels can be computed. The manufacturer's claimed reading sensitivity is of the order of 0.1 K, on absolute temperature accuracy of ± 1 K. The camera absolute calibration was done separately, using a heated blackened iron plate. In the present applications the interest is on the relative temperature variations $\Delta T$. Figure 3 illustrates two frames obtained during 10 December 2007 observations, one before and another during the 10:40 UT solar event exhibiting the displacement of the cross searching for the hottest pixel as the event intensity increases. $T_R$ and $T_H$ readings, respectively at left and right of the frames, give a $\Delta T$ increase of 1.2 K (i.e. corresponding to 107.8 and 106.6 °C, respectively).

We could not change the camera mode of functioning because we had no access to the proprietary software. This feature brings a serious problem when observing transients, since the searching cross moves into the position with maximum temperature, hides several pixels and makes analysis nearly impossible. To prevent the moving cross "interference" in solar observations using this camera we introduced a bright rim close to the solar limb, by slightly displacing the optical alignment, without affecting the solar image. In this condition the moving cross stays pointing to the bright rim, leaving the region of interest free from any cross searching interference. In this setting we cannot measure directly at the camera the temperature enhancements. Separate independent calibrations are necessary for accurate measurement of small and rapid changes.

The 10 December 2007 10:40 UT event was observed exceptionally without the bright rim at the solar limb, leaving the moving cross searching freely over the solar disk, and therefore allowing, for the first time, the direct temperature enhancement readings $\Delta T$ for each frame throughout the burst duration.

Mid-IR activity was occurring on AR 978, S08 E20, on the 10 December 2007 event, producing the spikes measured at the camera shown in Figure 4. Panel (a) indicates the temperature brightness differences $\Delta T$ for every successive frame readings (containing 4050 frames along 13.5 minutes of time). It exhibits a clear increase in the rate of $\Delta T$ excess temperature enhancements, ranging from 0.5 to 2.0 K. Panel (b) shows the same data time integrated over 20 seconds. Panel (c) shows the corresponding GOES B2.0-class burst.

A detailed morphology analysis of the sources on the active region producing the observed $\Delta T$ was possible for the phase prior to the main enhancement of the event, just before the moving cross "blurred" the area of the flaring region. It was done for the first 1270 frames corresponding to the time interval 10:39:35 – 10:43:49 UT. Figure 5 presents typical images sequences along 20 seconds. Examples of mid-IR flashes lasting a few seconds on the 4.6 x 4.6 arcmin frames containing the active region are shown in Figure 5(a). The isocontour levels 0.65, 0.70, 0.75, 0.80, 0.85, 0.90, 0.95, and 0.99 are in the same scale, with the level 1.0 corresponding to the intensity of the brightest observed pixel in the whole set of 1270 frames. Figure 5(b) shows the control neighboring solar quiescent region (dashed box in Figure 1), taken at the same times as the sequence shown in (a). The isocontours at the control quiescent region never exceeded the 0.1 level for the whole set of 1270 frames.

## 4. Discussion

The flux density at Earth for the short-lived mid-infrared flashes can be estimated using the Rayleigh-Jeans approximation valid for the 10 μm band (Phillips, 1987)

$$S = 2 k \Delta T/A \quad W\ m^{-2}\ Hz^{-1} \qquad (1)$$

where k is the Boltzmann constant and A the effective aperture area. Considering an aperture efficiency of 50% for the 0.105 m diameter mirror we obtain $A = 4.3 \times 10^{-3}\ m^2$. For $\Delta T = 0.5$-2.0 K, as read at the camera output (see Figures 3 and 4(a)), we get fluxes of 3-13 x $10^{-21}$ W $m^{-2}$ $Hz^{-1}$ (or 30-130 solar flux units). These values might be accurate to order of magnitude. The considerably larger fluxes previously reported (Melo *et al.*, 2006) were later found to be mistaken because the camera´s intrinsic temperature scale was not used and calibration was inadvertently done with the camera at saturated temperature scale (Melo *et al.*, 2009).

The fast mid-IR source temperature enhancements at the solar surface can be calculated referring to a temperature scale at the Sun. Gezari, Livingston, and Varosi (1999) have estimated about 1000 K the difference of temperature between dark sunspots and the photosphere. Adopting that scale, the temperature enhancements detected in the frames (such as in the example shown in Figure 5) are compared to the spot umbra minus the photosphere temperature measured in the same frames, giving temperature enhancements in the range of $\Delta T_b \sim$ 50-150 K for the fast mid-IR flashes at the Sun.

The flux density attributed to the flashes might be determined also from the well-known formula (Kraus, 1986)

$$S = 2 k \Delta T_s\ \Omega_s/\lambda^2 \qquad (2)$$

where $\Delta T_s$ and $\Omega_s$ are the real source temperature enhancement and solid angle, respectively, and $\lambda$ is the wavelength. To compare this flux to that one determined at the camera with equation (1) $\Delta T_s$ and $\Omega_s$ should be known. However we may tentatively assume that a typical flash enhancement at the solar surface $\Delta T_b \sim 100$ K $\sim \Delta T_s$ corresponds to a source occupying the whole field-of-view (25 arcsec), i.e. $\Omega_s \sim 2 \times 10^{-8}$ sr, producing a flux of about $6 \times 10^{-19}$ W m$^{-2}$ Hz$^{-1}$ (or 6000 solar flux units) according equation (2). The large discrepancy compared to the flux received at the camera suggests the need further of investigations related to the actual mid-IR flash source angular size and brightness.

The mid-IR flashing region was located on the solar disk using the Hinode XRT images (Kosugi *et al*., 2007; Golub *et al*., 2007), obtained at about the same time, as shown in Figure 6. In Figure 6(a) the mid-infrared frame on the active region is shown superimposed to the Hinode G band image obtained at 10:45:30 UT. Mid-IR spots A, B, and the fainter spot C shown in Figure 2 correspond well to the Hinode spots. The mid-IR diffraction limited 25 arcsec field-of-view is also indicated as circles. The same mid-infrared frame is superimposed to the dark background contrast Hinode image obtained at 10:45:34 UT in Figure 6(b). There is a qualitative suggestion that the mid-IR flashes occur on the bright soft X-ray patch in Figure 6(b).

## 5. Concluding Remarks

Ground-based high cadence solar observations at the 10 μm using a small aperture telescope proved to be an important tool to investigate transients associated to flares observed at other wavelength bands. Multiple small (arcseconds) and short-lived mid-IR flashes were characterized on active regions every time soft X-ray bursts are detected by GOES satellites. Mid-IR flashes were analyzed in detail for an event on 10 December 2007. Temperature enhancements measured at the camera correspond to flux densities of tenths of solar flux units. The frames analyzed of the active region exhibit temperature enhancements of order of 50-150 K when referred to a temperature scale of 1000 K between the spot umbra and the photosphere (Gezari, Livingston and Varosi, 1999) for sources unresolved within the 25 arcseconds FOV. The nature of the mid-IR flashes is unknown. They might be thermal responses to small scale instabilities related to the flaring phenomena or to rapid density fluctuations driven by magneto-hydrodynamic modulation of the propagation media opacity. The actual intrinsic temperature rise, however, only can be determined for resolved sources angular sizes. New observational mid-IR runs are planned to be carried out simultaneously with the high time resolution solar submillimeter telescope.

*Acknowledgments*
The authors gratefully acknowledge the helpful comments given by one anonymous referee which improved the paper presentation. This research was partially supported by Brazilian agencies FAPESP, CNPq, and Mackpesquisa, Argentina agency CONICET, and US agency AFOSR.

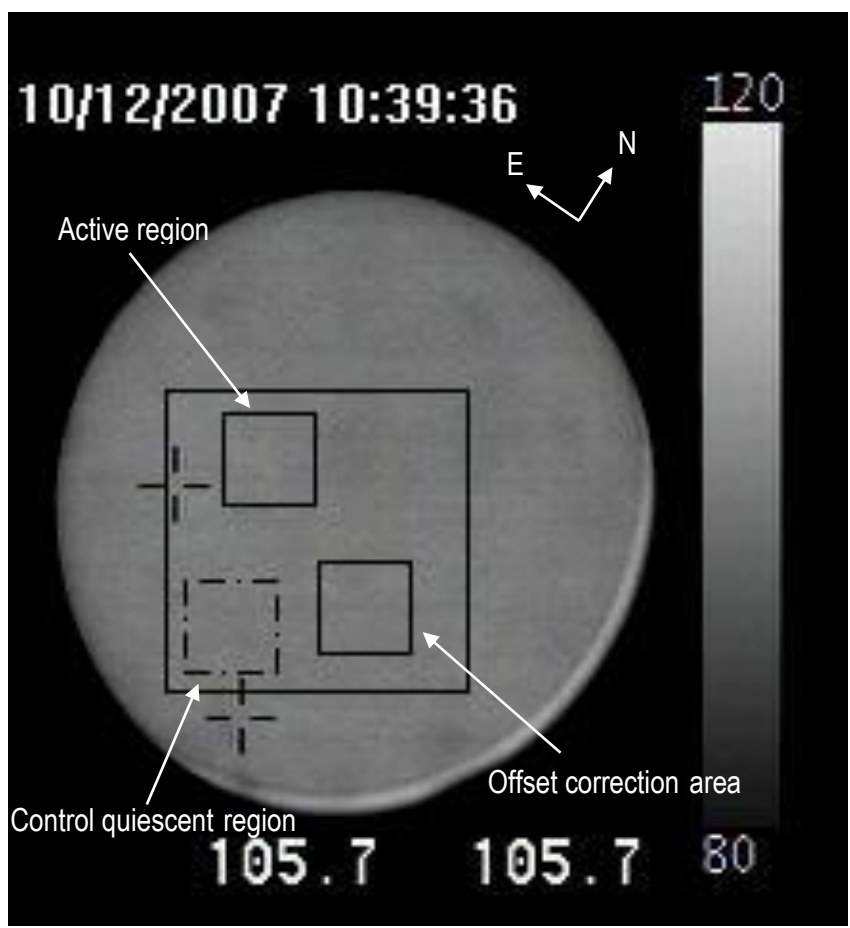

Figure 1 – Example of an original nearly full disk mid-infrared image captured during solar observations in 10 December 2007. The 80 x 80 pixels work area (14.7 x 14.7 arcmin), contains the upper left 25 x 25 pixels box (4.6 x 4.6 arcmin) where the active region is located. The lower right 25 x 25 pixels box is used for offset corrections. The dashed 25 x 25 pixels box is the control quiescent region.

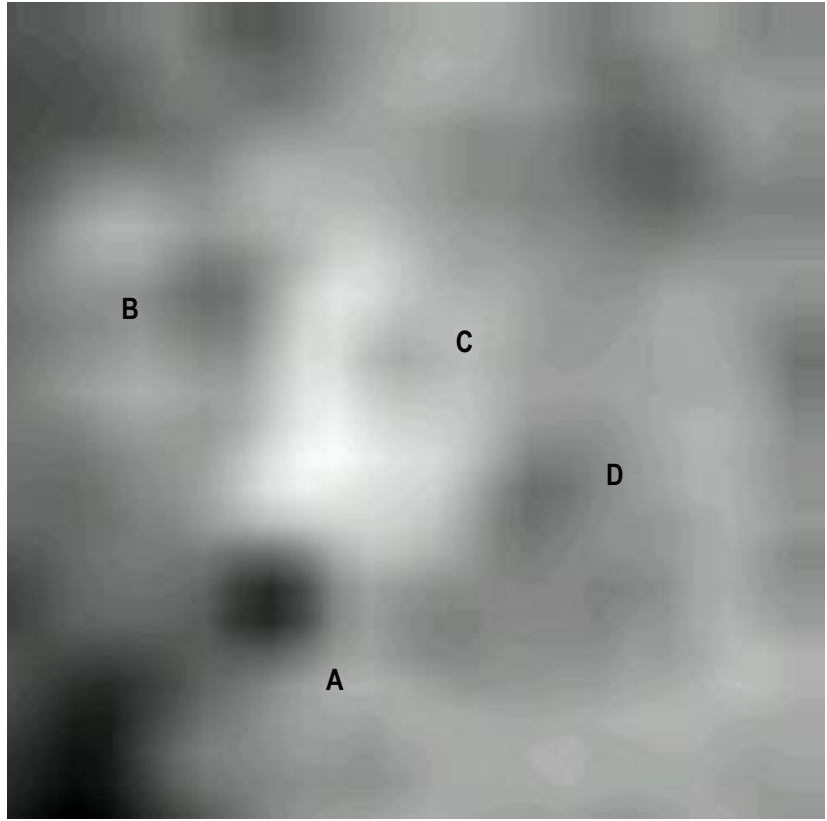

Figure 2 – Example of a resulting 25 x 25 pixels mid-infrared image obtained after the application of the sequence of image treatment procedures. It corresponds to an area of 4.6 x 4.6 arcmin containing the solar active region. Dark sunspots are labeled A and B. Spot C can be distinguished with less contrast. The feature D is due to dirt in the optical mirror. The frame is rotated clockwise by about 35° with respect to the solar N-S direction.

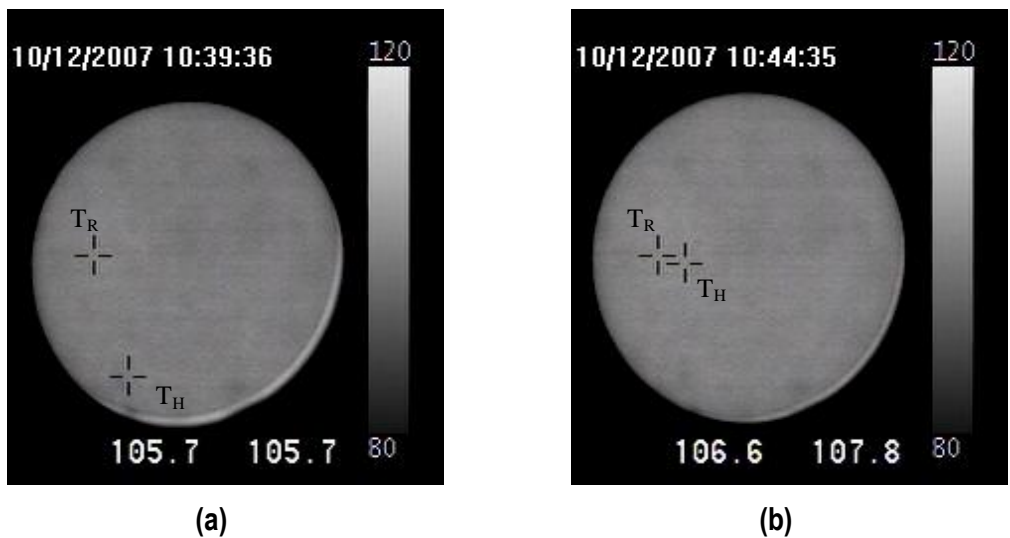

Figure 3 – Mid-infrared solar images obtained on 10 December 2007 before (a) and during (b) the 10:40 UT solar event. The cross displayed by the camera internal system reading the reference temperature ($T_R$) remains at a fixed position, while the moving cross indicates the temperature of the hottest pixel in the whole field ($T_H$). The difference $\Delta T$ between $T_R$ and $T_H$ values, respectively at left and right of the frames, can be computed. In this example we observe a $\Delta T$ increase of 1.2 K.

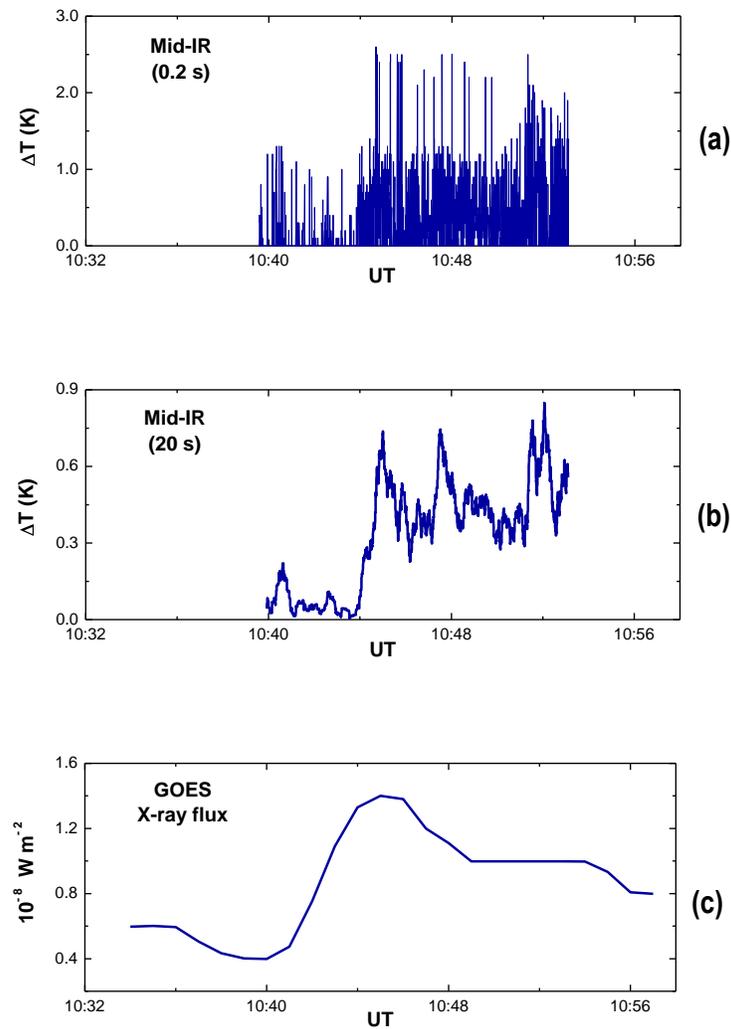

Figure 4 – (a) Temperature brightness difference $\Delta T$ between the searching cross defined hottest ($T_H$) temperatures and the fixed reference ($T_R$), obtained for every single frame recorded (4050 readings along 13.5 minutes of time). (b) Same data 20 seconds time integrated. (c) X-ray solar flux detected by GOES satellites during a B2.0-class event beginning at 10:40 UT (adapted from NOAA/SWPC, http://www.swpc.noaa.gov).

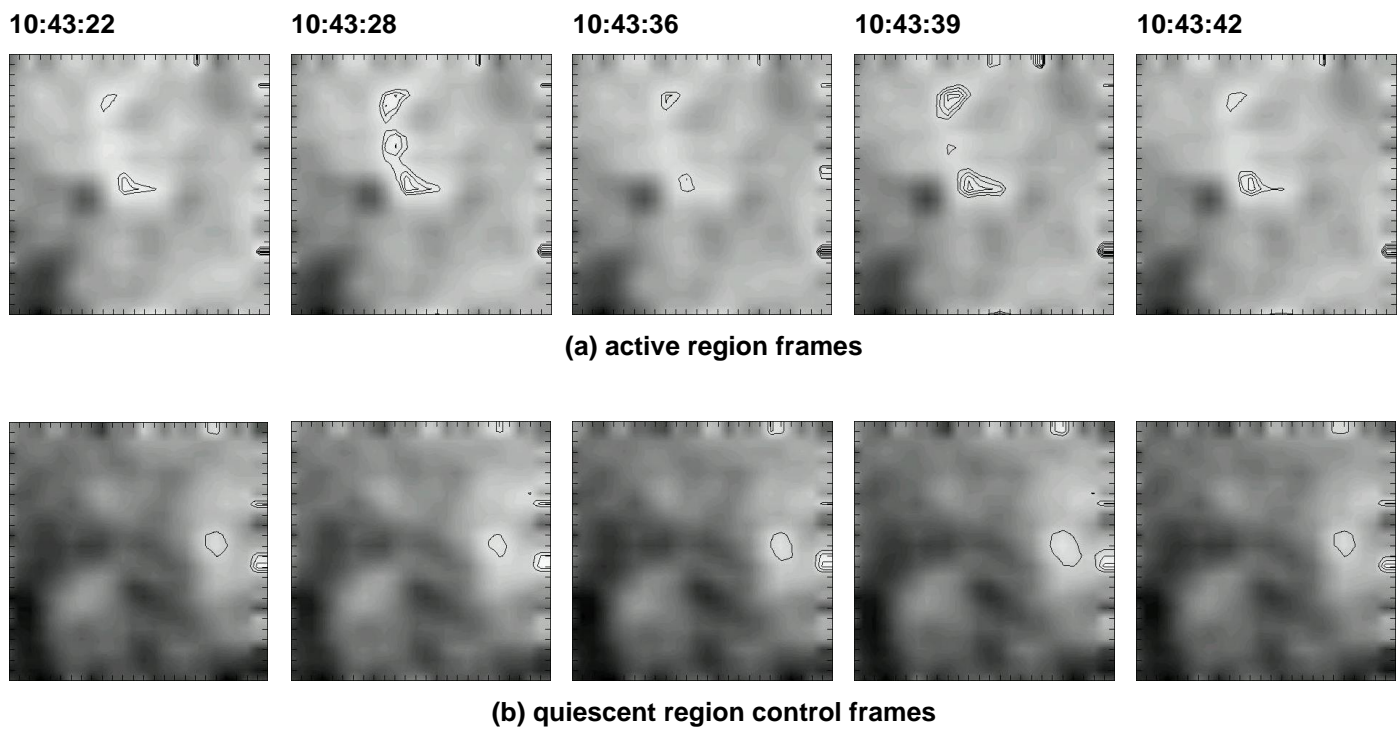

Figure 5 – Example of typical mid-infrared image sequences obtained during a total time interval of 20 seconds, on 10 December 2007. Sequence (a) shows fast mid-IR flashes, with time scales of seconds, at different locations in the active region. Sequence (b) shows control frames, obtained at the same times, on a neighboring solar quiescent region (dashed box in Figure 1). The isocontour levels are on the same scale, with the level 1.0 corresponding to the intensity of the brightest observed pixel in the whole set of 1270 frames. In (a), the displayed levels are 0.65, 0.70, 0.75, 0.80, 0.85, 0.90, 0.95, and 0.99. At the control quiescent region (b) features never exceeded the 0.1 level.

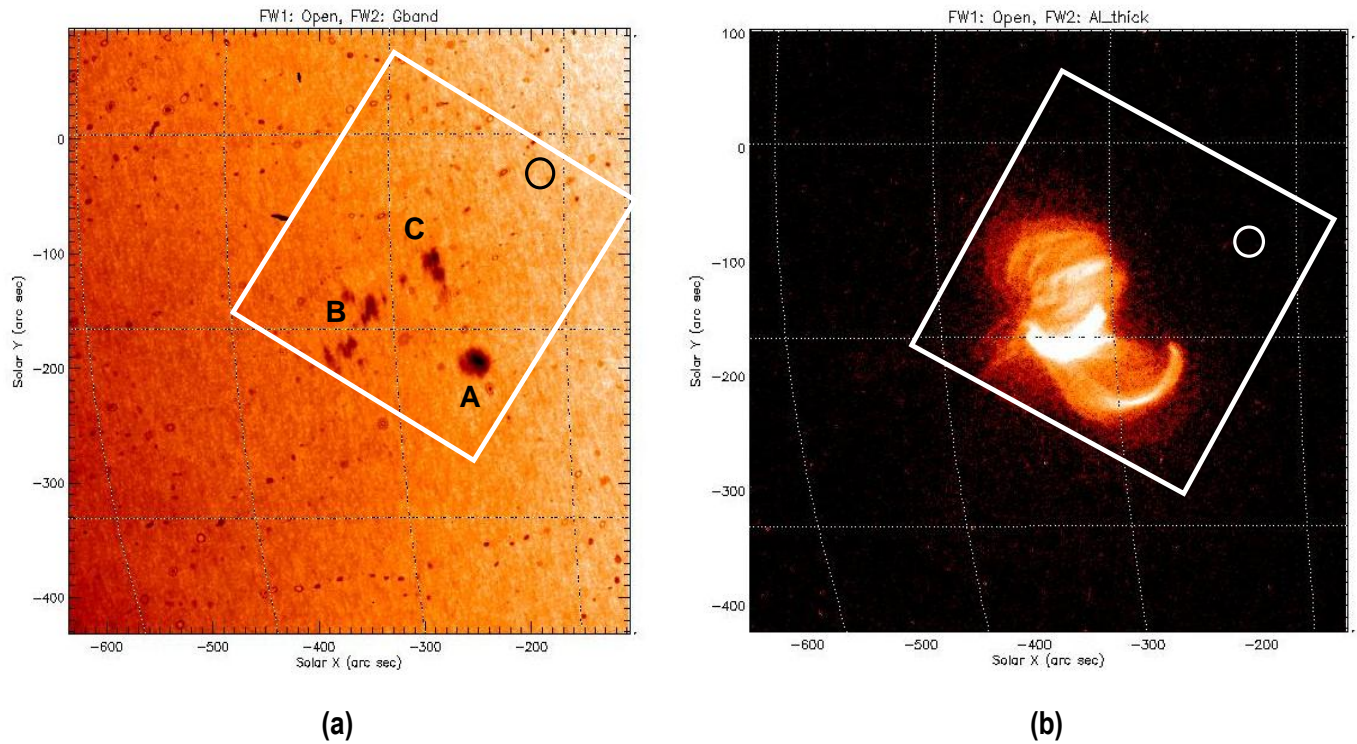

(a)  (b)

Figure 6 – The mid-infrared active region frames were located in Hinode XRT images. The position angle of the IR frames (white box) is rotated clockwise by about 35$^o$ with respect to the solar N-S direction. (a) Mid-IR frame with filter G band image, obtained at 10:45:30 UT, and (b) Al_thick image, obtained at 10:45:34 UT. Spots A, B, and C can be compared to the mid-infrared spots shown in Figure 2. The circles indicate the 25 arcsec mid-infrared field-of-view. (Hinode quicklook images, http://sdc.uio.no/search/API).